# The dynamic electric polarizability of a particle bound by a double delta potential

M. A. Maize[a)] and J. J. Smetanka[b)]
*Department of Physics, Saint Vincent College, 300 Fraser Purchase Road, Latrobe, Pennsylvania 15650*

In this paper we derive an expression for the dynamic electric polarizability of a particle bound by a double delta potential for frequencies below and above the absolute value of the particle's ground state energy. The derived expression will be used to study some of the fundamental features of the system and its representation of real systems. In addition we derive a general expression of the dynamic electric polarizability for a potential of multi-delta functions.

## I: INTRODUCTION

The study of the electric polarizability, α, has been significant in many areas of physics. Calculation of α has played an important role in educating us about the electromagnetic properties of atoms[1], the internal structure of the nucleons[2, 3], and the basic properties of solids[4], to name a few examples. Certainly, the study of the static electric polarizability, α(0), has received more attention than the study of the dynamic polarizability, α(ω), at least at the level of a fundamental treatment[5]. From an educational point of view, determination of α(ω) gives an excellent example of the application of perturbation theory to the interaction of systems with time-dependent external sources. Second, α(ω) provides us with great knowledge of the dispersive electromagnetic properties of atoms based on the study of the linear response of the atomic charge to an external oscillating electric field of various frequency, ω[1, 6]. Third, we can obtain the static electric polarizability α(0), by taking the static limit of α(ω). These are but a small fraction of the reasons we believe that the study of α(ω) is valuable to the basic learning of quantum mechanics and through its application enhancing our knowledge in many fields of physics and related disciplines.

Based on our fundamental knowledge of electromagnetism, application of an external electric field to a charge system can lead to charge displacement. The linear response of the charge displacement (distortion) is expressed in terms of what we define as the induced electric dipole. The induced electric dipole is proportional to the applied field and the proportionality is given by the electric polarizability. To obtain the quantum α(ω), we find the average of the electric dipole operator, and divide it by the external time-dependent electric field, ε(t)[5]. For the rest of the paper, the electric polarizability will always be in reference to the quantum mechanical problem.

The conventional perturbation method used to calculate either the static or dynamic electric polarizability involves an infinite sum or integral that contains all possible states allowed by the electric dipole transition[7]. Some of these states, for example, the scattering states, can be very difficult or impossible to obtain in a large number of cases[7]. In this work, the only states we are going to use in our calculation are the ground state of the unperturbed system and the complete set of the free particle states $|k\rangle \left( |k\rangle = \frac{1}{\sqrt{2\pi}} e^{ikx} \right)$.

The double delta potential has been used extensively for a variety of purposes[8-10]. Most notably, this potential has been employed to demonstrate the quantum features of a system that resembles the motion of an electron in the field of two nuclei[8]. A calculation of the eigenstates and eigenfunctions of the bound system of the double delta potential with comparison to the results for actual molecules had been reported by Lapidus[9]. Most recently, Bonfim and Griffiths considered the example of the double delta potential in presenting their approach to obtain closed form expressions for the bound-state energies in the one-dimensional problem[10].

In calculating α(ω), we study the perturbative interaction between a charged particle, which is initially in the ground state of the system, and an applied time-dependent electric field[5]. As reported in Refs. 8-10, the double delta potential allows for one or two bound (negative energy) states. We will do our calculations for both cases while staying within the "non-degenerate" conditions[10]. In section II we give a review of the model and the complete expression of the ground-state wavefunction. In section III we review the case of the single delta[5], and demonstrate how our technique can be used to determine α(ω) for the double delta potential and establish a general expression α(ω) for a multi-delta potential. In section IV, we present our numerical results with analysis. Finally, we give concluding remarks.

## II: THE WAVE EQUATION AND THE GROUND-STATE WAVE FUNCTION

The unperturbed Hamiltonian $H_0$ is given by

$$H_0 = \frac{-\hbar^2}{2m}\frac{\partial^2}{\partial x^2} - g\left[\delta(x+a)+\delta(x-a)\right], \tag{1}$$

where g is the strength of the potential and m is the mass of the particle occupying the ground state. Now let $p = \frac{2mga}{\hbar^2}$. For p < 1, the solution of the wave equation $H_0\Psi = E\Psi$ will produce one bound (negative energy) state and a continuum of unbound states. The bound state, we will call $\Psi_0$ is given by[8-10]:

$$\Psi_0\left(|x|>a\right) = N(k_0a)Cosh(k_0a)e^{-k_0(|x|-a)}, \tag{2a}$$

and

$$\Psi_0\left(|x|<a\right) = N(k_0a)Cosh(k_0x), \tag{2b}$$

where $N(k_0a) = \sqrt{\frac{2k_0}{(e^{2k_0a}+2k_0a+1)}}$, $k_0$ is related to the bound state energy, $E_0$, by:

$$E_0 = -\frac{\hbar^2k_0^2}{2m}, \tag{3}$$

with $k_0$ and $g$ related by the equation:

$$\frac{2mg}{\hbar^2} = k_0[1+Tanh(k_0a)]. \tag{4}$$

For $p \geq 1$, a second bound (negative energy) state will satisfy Eq. (1) in addition to $\Psi_0$ and a continuum of unbound states. As long as p is not close to or greater than 5, $\Psi_0$ is the lowest energy state of the system. For p close to or larger than 5, the eigenenergy of state $\Psi_0$ will be very close to the eigenenergy of the second bound state and we will be in the region of double degeneracy[10]. For our calculation of α(ω) in this paper, we consider only the non-degenerate case, $(p<5)$.

## III: THE CALCULATION OF α(ω)

To determine α(ω), we study the interaction between a particle of charge q that is initially in the ground state of the system and an external time-dependent electric field of variable frequency ω[5]. Taking the electric field to be parallel to the x-axis, the basic expression of α(ω) is[5]:

$$\alpha(\omega) = -q^2\sum_n\left[\frac{\langle\Psi_0|x|\Psi_n\rangle\langle\Psi_n|x|\Psi_0\rangle}{\hbar\omega - E_{n0}+i\mu} - \frac{\langle\Psi_0|x|\Psi_n\rangle\langle\Psi_n|x|\Psi_0\rangle}{\hbar\omega + E_{n0}+i\mu}\right] \tag{5}$$

Here $qx$ is the dipole operator and $E_{n0} = E_n - E_0$, where $E_n$ is the energy of the state $\Psi_n$. The functions $\Psi_n$ represent a complete set of eigenfunctions of the Hamiltonian $H_0$[5].

To calculate the matrix elements appearing in Eqs. (5), we use the bound state wave function of Eqs. (2a) and (2b) and the completeness of the free particle states $|k\rangle$ ($|k\rangle = \frac{1}{\sqrt{2\pi}} e^{ikx}$). Due to this we write $\alpha(\omega)$ as:

$$\alpha(\omega) = -q^2 \left[ \int \langle \Psi_0 |x| k' \rangle \langle k' | G(+\omega) | k \rangle \langle k |x| \Psi_0 \rangle dk dk' + \int \langle \Psi_0 |x| k' \rangle \langle k' | G(-\omega) | k \rangle \langle k |x| \Psi_0 \rangle dk dk' \right], \tag{6}$$

where $G(\pm\omega)$ is defined by:

$$G(\pm\omega) = G_0(\pm\omega) + G_0(\pm\omega) V(x) G(\pm\omega), \tag{7}$$

with $V(x)$ being the binding potential. $G(\pm\omega)$ and $G_0(\pm\omega)$ are given in terms of $\hat{T}$, $H_0$, $\hbar\omega$ and $E_0$ by:

$$G(\pm\omega) = \frac{1}{E_0 \pm \hbar\omega - H_0}, \tag{8}$$

$$G_0(\pm\omega) = \frac{1}{E_0 \pm \hbar\omega - \hat{T}}, \tag{9}$$

where $\hat{T} = -\frac{1}{2}\frac{d^2}{dx^2}$. To calculate $\alpha(\omega)$ (Eq. (6)), we need to calculate the matrix elements $\langle \psi_0 |x| k' \rangle \langle k' | G(\pm\omega) | k \rangle \langle k |x| \psi_0 \rangle$. Calculations of the matrix elements $\langle \psi_0 |x| k' \rangle$ and $\langle k |x| \psi_0 \rangle$ for either the double or the single delta can be done with a simple one-dimensional integration. Now to calculate $\langle k' | G(\pm\omega) | k \rangle$ (Eq. (7)), we first obtain for $\langle k' | G_0(\pm\omega) | k \rangle$ the following result:

$$\langle k' | G_0(\pm\omega) | k \rangle = G_{0k}(\pm\omega) \delta(k - k'), \tag{10}$$

where $G_{0k}(\pm\omega) = \frac{1}{E_0 \pm \hbar\omega - \frac{\hbar^2 k^2}{2m}}$. With the results of Eq. (9), $\langle k' | G(\pm\omega) | k \rangle$ can then be written as:

$$\langle k' | G(\pm\omega) | k \rangle = G_{0k}(\pm\omega)\delta(k - k') + \int \langle k' | G_0(\pm\omega) | k_1 \rangle \langle k_1 | V(x) | k_2 \rangle \langle k_2 | G(\pm\omega) | k \rangle dk_1 dk_2, \tag{11}$$

with both $|k_1\rangle$ and $|k_2\rangle$ are complete sets of free particle states. Now for the single delta $V(x) = -g\delta(x)$ and by substitution this potential in Eq. (11) we get

$$\langle k' | G(\pm\omega) | k \rangle = G_{0k}(\pm\omega)\delta(k - k') - \frac{g}{2\pi} G_{0k'}(\pm\omega) A(\pm\omega, k), \tag{12}$$

$G_{0k'}(\pm\omega)$ is obtained by replacing k by k′ in the expression of $G_{0k}(\pm\omega)$ and $A(\pm\omega,k)$ is given by

$$A(\pm\omega, k) = \int \langle k_2 | G(\pm\omega) | k \rangle dk_2. \tag{13}$$

For the more general case of the double delta potential $V(x) = -[g_1\delta(x+a) + g_2\delta(x-a)]$ and by substituting with this potential in Eq. (11), we get

$$\langle k' | G(\pm\omega) | k \rangle = G_{0k}(\pm\omega)\delta(k - k') - \frac{G_{0k'}(\pm\omega)}{2\pi} \left[ g_1 A_1(\pm\omega, k) e^{ik'a} + g_2 A_2(\pm\omega, k) e^{-ik'a} \right]. \tag{14}$$

where

$$A_1(\pm\omega, k) = \int e^{-ik_2 a} \langle k_2 | G(\pm\omega) | k \rangle dk_2, \tag{15a}$$

and

$$A_2(\pm\omega, k) = \int e^{ik_2 a} \langle k_2 | G(\pm\omega) | k \rangle dk_2. \tag{15b}$$

By comparing Eqs. (12) and (14), we can see that we can write $\langle k'|G(\pm\omega)|k\rangle$ for any number of deltas. If we keep adding delta functions at +2a with strength $g_3$ and at -2a with strength $g_4$ and so on then $\langle k'|G(\pm\omega)|k\rangle$ is given by

$$\langle k'|G(\pm\omega)|k\rangle = G_{0k}(\pm\omega)\delta(k-k') - \frac{G_{0k'}(\pm\omega)}{2\pi}\begin{bmatrix} g_1 A_1(\pm\omega,k)e^{ik'a} + g_2 A_2(\pm\omega,k)e^{-ik'a} \\ + g_3 A_3(\pm\omega,k)e^{2ik'a} + g_4 A_4(\pm\omega,k)e^{-2ik'a} + \cdots \end{bmatrix} \tag{16}$$

Substituting Eq. (16) in Eq. (11) will then give us the generalized expression for α(ω) for any number of deltas. Now if we integrate both sides of Eq. (12) over *k'* we can solve for *A*. If we multiply both sides of Eq. (14) by $e^{ik'a}$ integrate over *k'*, we can write $A_2$ in terms of $A_1$ and we have both coefficients. Getting the general expression of the coefficients for any number of terms (Eq. (16)) is then a straightforward process. Now we have the expression for $\langle k'|G(\pm\omega)|k\rangle$, we can substitute it in Eq. (6) to get α(ω). We will give the expressions for *A*, $A_1$ and $A_2$ in appendix A. Complete expressions for $\alpha(\omega < \omega_B)$ and for $\alpha(\omega > \omega_B)$ where $\omega_B = \frac{E_B}{\hbar}$ with
$E_B = -E_0$, will be given in a web site (in appendices B-G). Finally, for the remainder of the paper we set
$g_1 = g_2 = g$.

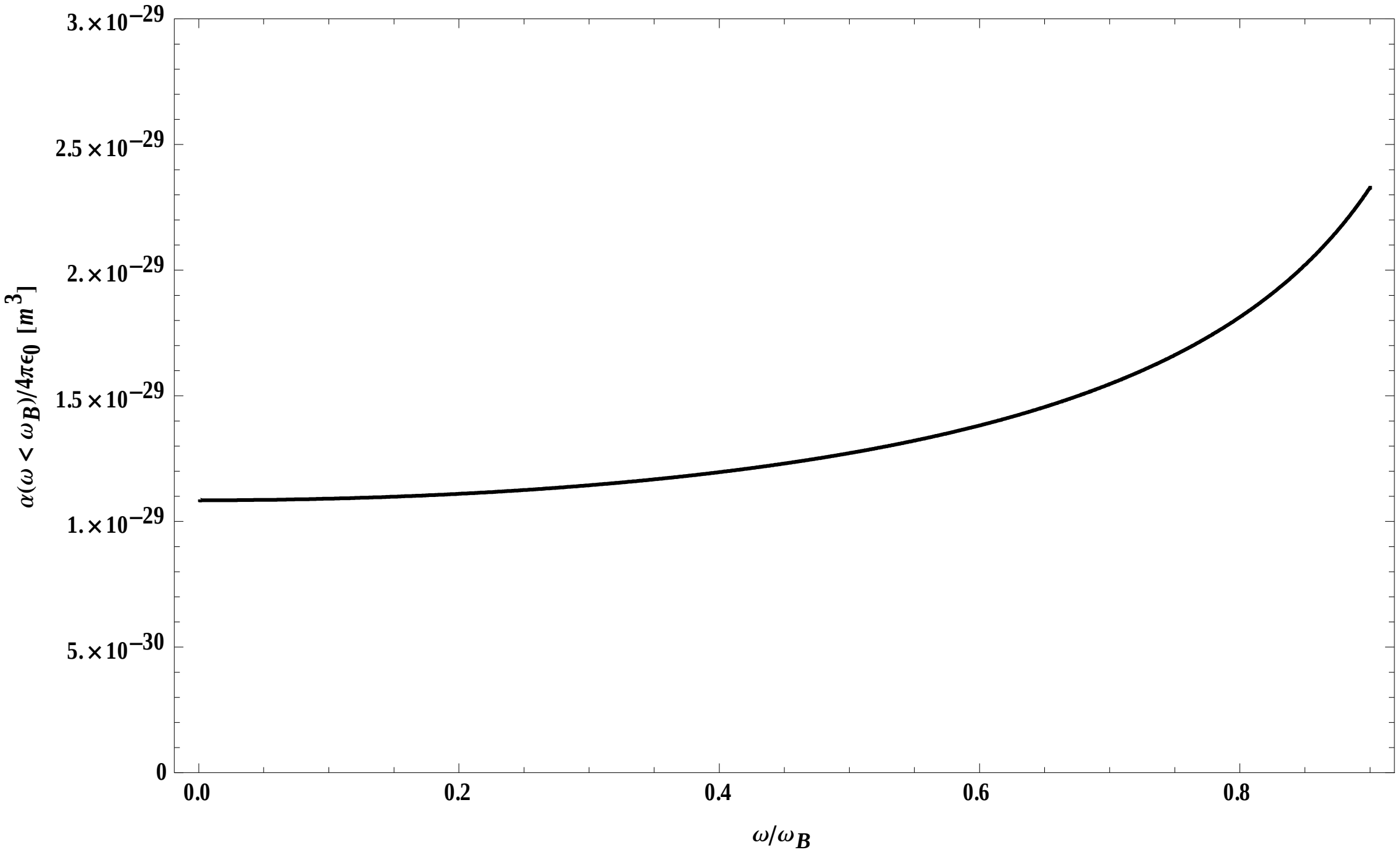

*Fig. 1a. Plot of* $\frac{\alpha(\omega<\omega)}{4\pi\varepsilon_0}$ *in* $m^3$ *vs.* $\frac{\omega}{\omega_B}$ *where* $\omega_B = \frac{E_B}{\hbar}$*, 2a = 1Å, and p = 0.5.*

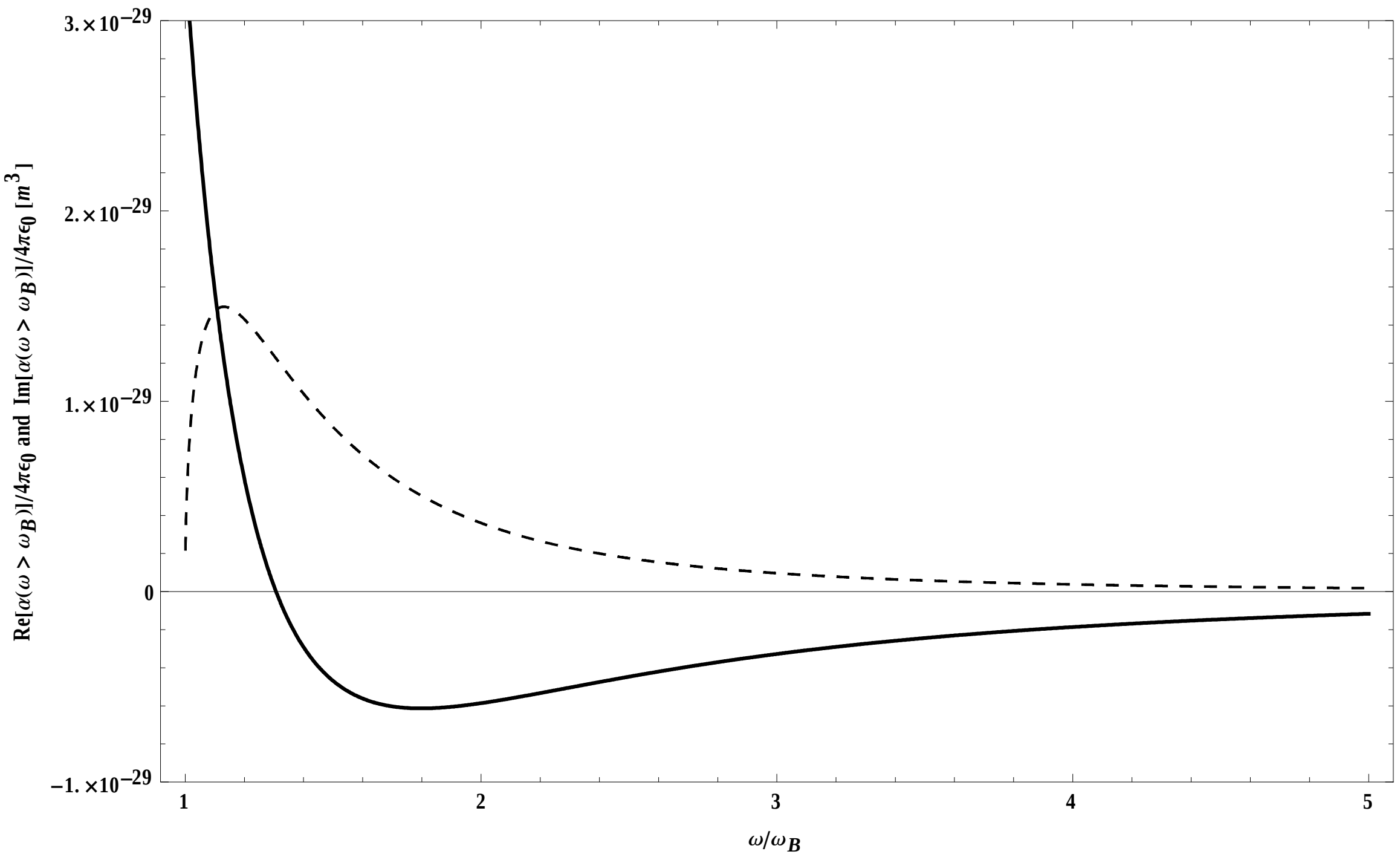

*Fig 1b. Plot of* $\frac{\text{Re}[\alpha(\omega > \omega_B)]}{4\pi\varepsilon_0}$ *(solid line) and* $\frac{\text{Im}[\alpha(\omega > \omega_B)]}{4\pi\varepsilon_0}$ *(dashed line) in* $m^3$ *vs.* $\frac{\omega}{\omega_B}$ *for 2a = 1Å, and p = 0.5.*

## IV: NUMERICAL RESULTS AND ANALYSIS

In all of our graphs, the vertical axis represents $\alpha(\omega)$ as given by Eq. (6) multiplied by $\frac{1}{4\pi\varepsilon_0}\ (9.0X10^9\ \frac{Nm^2}{C^2})$ so that the final unit is in $m^3$. We do so because the polarizability is usually given in units of distance cubed as introduced in undergraduate textbooks.[11] In Figs.1a and 1b, the dynamic electric polarizability displays the same characteristics as the case of the single delta[5]. Such similarity is not surprising since for p < 1 there is one bound (negative energy) state and a continuum of unbound states which is the same structure as in the case of the single delta. It is also important to observe that the threshold of the photoelectric effect takes place at $\hbar\omega = E_B$ (Figs. 1b and 2b). We know from our studies of the photoelectric effect that when $\hbar\omega \geq E_B$ ($E_B$ here represents the work function) the total absorption of a photon of energy $\hbar\omega$ can take place. Such absorption leads to the existence of the imaginary part.

In figure 2a we can see that the polarizability blows up when $\omega$ is close to $0.75\omega_B$ (numerical calculations give a more precise value of $0.7529\omega_B$). This behavior has been reported before in the study of the atomic dynamic electric polarizability and examples can be found in Refs. 1, 12, and 13. But why does $\alpha(\omega)$ for $\omega < \omega_B$ ($\alpha(\omega < \omega_B)$) blow up in the case of p = 1.5 and not in the case of p = 0.5? For p ≥ 1, the unperturbed system contains two bound states[8-10]. The blow up in $\alpha(\omega < \omega_B)$ takes place at an $\omega$ where $\hbar\omega$ is equal to the energy difference between the two bound states. Since the only bound state we used in our calculation is $\Psi_0$, our expression for the dynamic polarizability is then responsible for detecting the existence of a second bound state for the case of p = 1.5. Let us define $h\omega_1$ to be the absolute value of the eigenenergy of the second bound state. Now the resonant frequency at $\omega = 0.7529\omega_B$ should be given by $h\omega = E_B - h\omega_1$ resulting in $\omega_1 = 0.2471\omega_B$. Using the detailed expression derived for the bound states[8-10], we get the same value for $\omega_1$. From a pedagogical point of view, this gives a clear illustration of how the study of polarizability can be used to study the structure of a bound system.

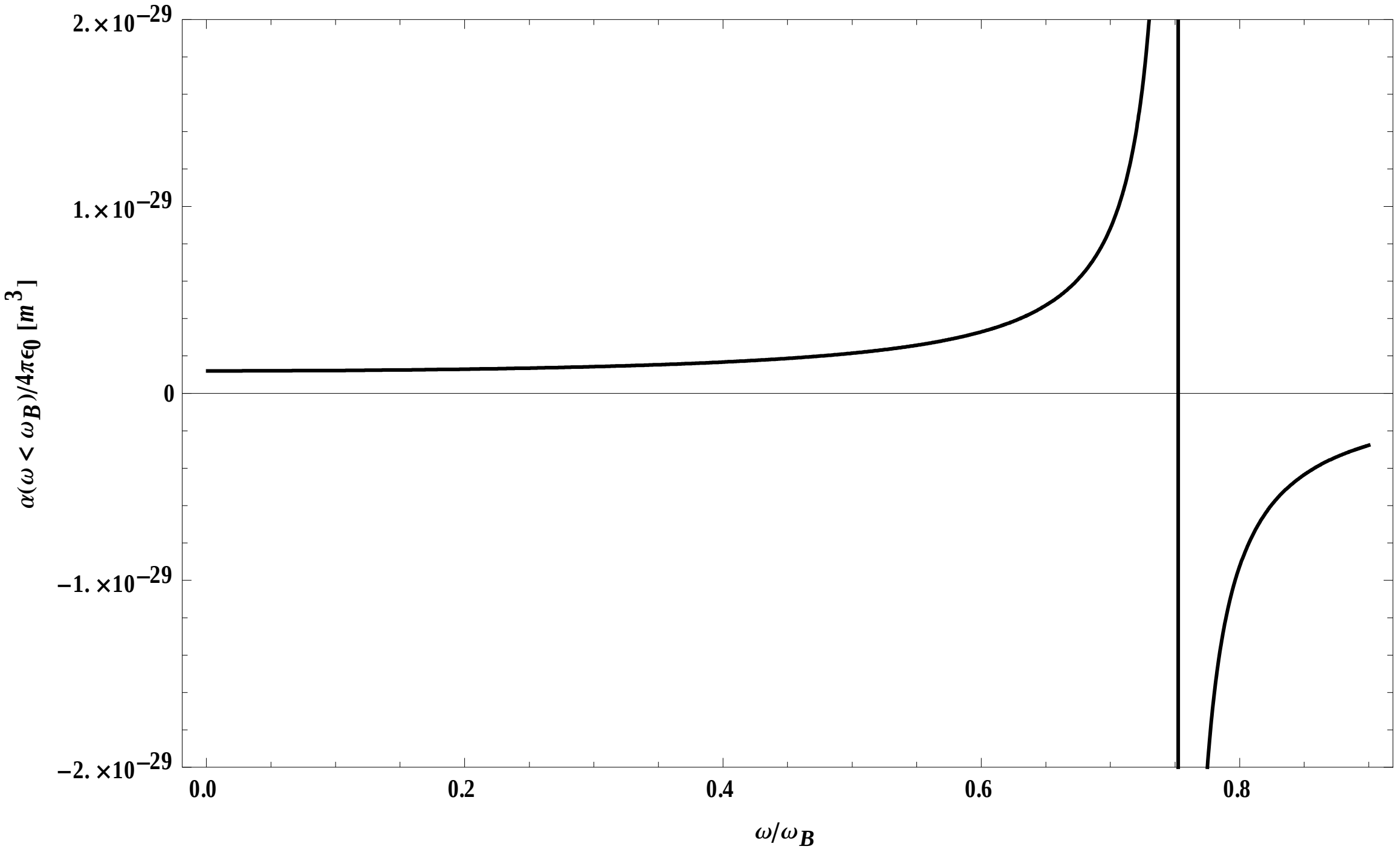

*Fig. 2a. Plot of* $\dfrac{\alpha(\omega < \omega)}{4\pi\varepsilon_0}$ *in* $m^3$ *vs.* $\dfrac{\omega}{\omega_B}$ *where 2a = 1Å, and p = 1.5.*

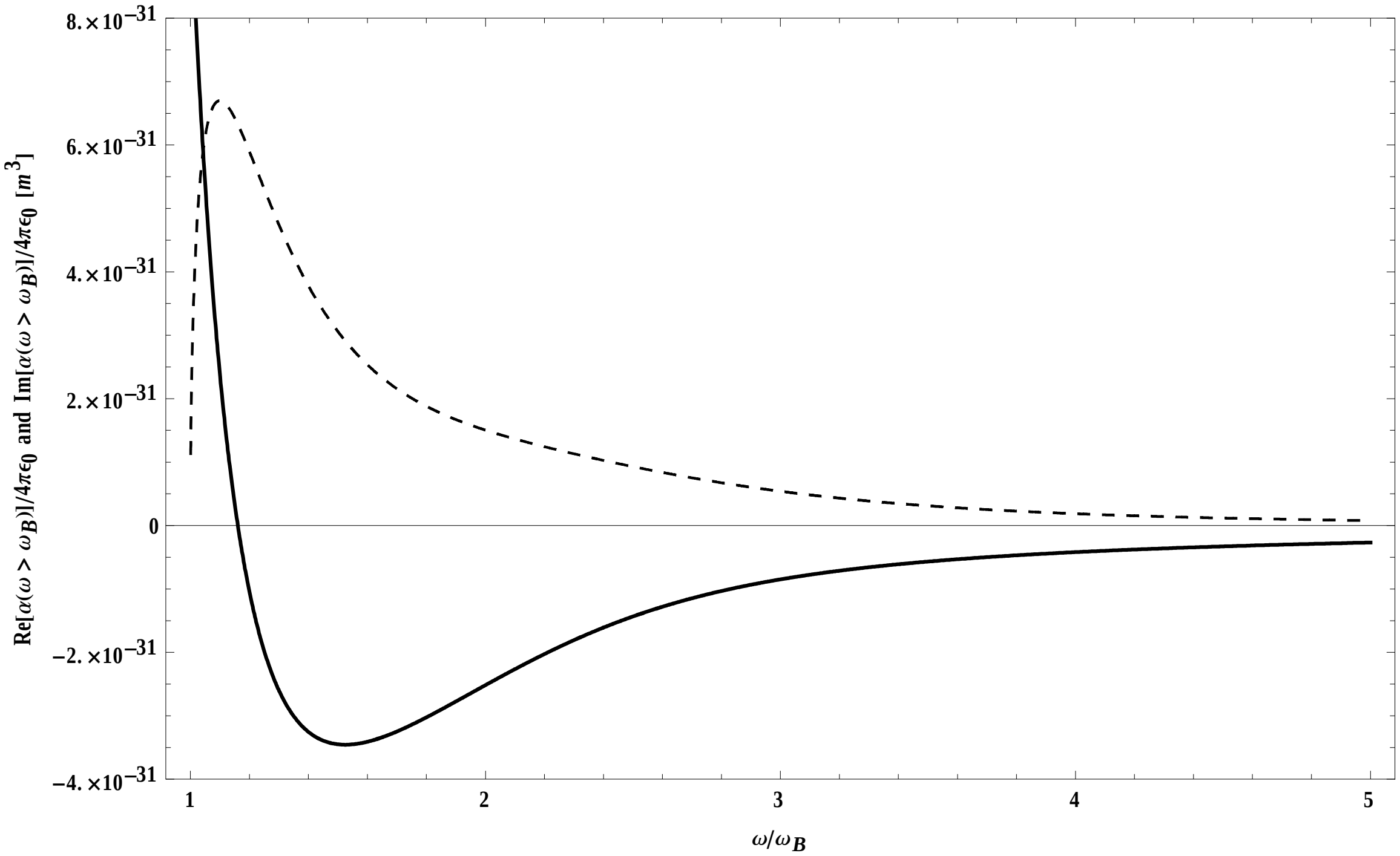

*Fig 2b. Plot of* $\dfrac{\mathrm{Re}[\alpha(\omega > \omega_B)]}{4\pi\varepsilon_0}$ *(solid line) and* $\dfrac{\mathrm{Im}[\alpha(\omega > \omega_B)]}{4\pi\varepsilon_0}$ *(dashed line) in* $m^3$ *vs.* $\dfrac{\omega}{\omega_B}$ *for 2a = 1Å and p = 1.5.*

In comparing Figs. 1b and 2b, we can see that the increase of p leads to a higher threshold frequency for the photoelectric effect ($\omega_B$ is higher for p = 1.5). This is because increasing p leads to stronger attraction.

The static polarizability $\alpha(0)$ can be obtained by setting $\omega = 0$ in our expression for $\alpha(\omega)$ (Eq. (6)) and performing a set of simple integration or by taking $\omega \to 0$ in the final expression of $\alpha(\omega)$ (Appendix C). When we set q = e, m = $m_e$, 2a = 0.74Å and p = 1.22 in our expression for $\alpha(0)$, we obtain 4.68 X $10^{-31}$ $m^3$. The acceptable value for the static polarizability of the $H_2$ ion is 4.69 X $10^{-31}$ $m^3$.[14] This agreement demonstrates that simple models can be used within reason to study realistic systems. To study the relation between the static polarizability and the structure of the system, we use the basic expression for $\alpha(0)$ which can be obtained by setting $\omega$=0 in Eq. (5):

$$\alpha(0) = 2q^2 \sum_n \frac{\left|\left\langle \Psi_n \left| x \right| \Psi_0 \right\rangle\right|^2}{E_n - E_0}, \tag{17}$$

where the summation over n does not include the state $\Psi_0$. Referring to Fig. 3, $\alpha(0)$ at $p$ = 1 is approximately 1/5 its value at $p$ = 0.5. For $p$ < 1, the states $\Psi_n$'s belong to the continuum. Increasing $p$ does increase $g$ and this will lead to the increase in the energy gap $E_n$ - $E_0$. The increase in the energy gap is then instrumental in $\alpha(0)$ decreasing while $p$ increases. $\alpha(0)$ at $p$ = 1.5 is approximately ½ its value at $p$ = 1.0. From $p$ = 1.5 to $p$ = 2.5, $\alpha(0)$ is about constant and then there is a modest increase in its value from $p$ = 2.5 to close to the end of the "non-degenerate" region. The behavior of $\alpha(0)$ beyond $p$ = 1.0 does not then continue the precipitous drop which occurs for $p$ < 1.0. As we discussed earlier, for $p \geq 1$, the system of states in our problem will contain two bound (negative energy) states in addition to the continuum. Due to this the polarization of the system will come from the transitions $\Psi_0$ to second bound state and $\Psi_0$ to continuum states. So we can conclude that the transitions $\Psi_0$ to the second bound state is what stopped the rapid decrease in $\alpha(0)$. In addition, there is then a decrease in the energy gap between the two bound states which leads to the modest increase of $\alpha(0)$ shown in Fig. 4 for $p$ > 2.5.

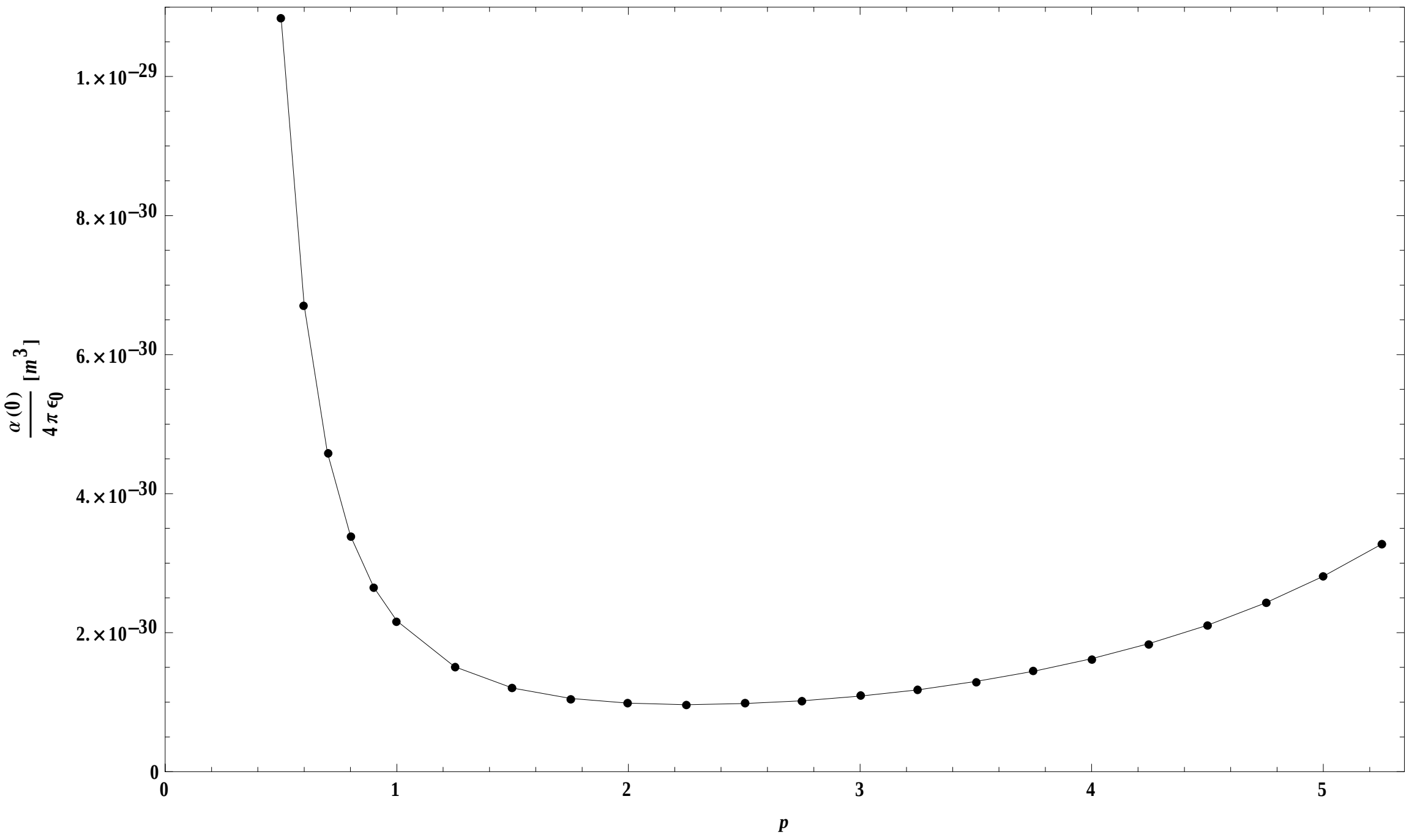

*Fig. 3. Plot of* $\dfrac{\alpha(0)}{4\pi\varepsilon_0}$ *in* $m^3$ *vs. p for systems with 2a = 1Å.*

## V: CONCLUSIONS

In this paper we demonstrate the effectiveness of relying on $\Psi_0$ and the complete set of free particle states $\left|k\right\rangle$, to derive the final expression for $\alpha(\omega)$ in the case of a particle bound by a double-delta potential and establish a general expression for $\alpha(\omega)$ in the case of the multi-delta potential. We believe that the work we present here is a modest addition to our learning of applications in the area of perturbing charged systems with time-dependent fields of variable frequency. Second, we demonstrate the relationship between the electromagnetic properties of a system and its structure. Finally, this problem can aid in preparing physics students toward doing more advanced work in many areas of theoretical physics.

## APPENDIX A: EXPRESSIONS FOR $A(\pm\omega,k)$, $A_1(\pm\omega,k)$, AND $A_2(\pm\omega,k)$

$$A(\pm\omega,k) = \frac{G_{0k}(\pm\omega)}{I(\pm\omega,k)}, \tag{A1}$$

where

$$I(\pm\omega,k) = 1 + \frac{g}{2\pi}\int G_{0k'}(\pm\omega)dk' \tag{A2}$$

$E_0$ in the expression of $G_{0k'}(\pm\omega)$ refers to the ground state energy of the single delta.[5]

$$A_1(\pm\omega,k) = G_{0k}(\pm\omega)\left[\frac{Cos(ka)}{I_1(\pm\omega,k)} - \frac{iSin(ka)}{I_2(\pm\omega,k)}\right], \tag{A3}$$

where

$$I_1(\pm\omega,k) = 1 + \frac{g}{2\pi}\int G_{0k'}(\pm\omega)(1+Cos(2k'a))dk', \tag{A4}$$

$$I_2(\pm\omega,k) = 1 + \frac{g}{2\pi}\int G_{0k'}(\pm\omega)(1-Cos(2k'a))dk', \tag{A5}$$

$$A_2(\pm\omega,k) = G_{0k}(\pm\omega)\left[\frac{e^{ika} - I_3(\pm\omega,k)A(\pm\omega,k)}{I_1(\pm\omega,k) - I_3(\pm\omega,k)}\right], \tag{A6}$$

where

$$I_3(\pm\omega,k) = \frac{g}{2\pi}\int G_{0k'}(\pm\omega)Cos(2k'a)dk' \tag{A7}$$

## APPENDIX B: $\alpha_-(\omega)$ AND DEFINING THE PARTS OF $\alpha_+(\omega)$

Since the derived expression of $\alpha(\omega)$ is a long one we write $\alpha(\omega) = \alpha_+(\omega) + \alpha_-(\omega)$. We choose to write $\alpha_+(\omega)$ in the form of a collection of terms and then we give a detailed expression for each term. $\alpha_+(\omega)$ then is given by:

$$\alpha_+(\omega) = \alpha_{+1}(\omega) + \alpha_{+2}(\omega). \tag{B1}$$

The first term on the right hand side of Eq. (B1) corresponds to replacing $G(+\omega)$ in Eq. (6) by $G_0(+\omega)$ and the second term corresponds to replacing $G(+\omega)$ in Eq. (6) by $G_0(+\omega)V(x)\,G(+\omega)$. We will give the detailed expressions for $\alpha_{+1}(\omega)$ and $\alpha_{+2}(\omega)$ in the case of $\omega < \omega_B$ and $\omega > \omega_B$. According to this $\alpha_+(\omega)$ will then be given as:

$$\alpha_+(\omega < \omega_B) = \alpha_{+1}(\omega < \omega_B) + \alpha_{+2}(\omega < \omega_B), \tag{B2}$$

and

$$\alpha_+(\omega > \omega_B) = \alpha_{+1}(\omega > \omega_B) + \alpha_{+2}(\omega > \omega_B). \tag{B3}$$

$\alpha_-(\omega)$ is obtained by replacing ω by -ω in Eq. (A2). $\alpha(\omega)$ can be written for $\omega < \omega_B$ and $\omega > \omega_B$ as:

$$\alpha(\omega < \omega_B) = \alpha_+(\omega < \omega_B) + \alpha_-(\omega), \tag{B4}$$

$$\alpha(\omega > \omega_B) = \alpha_+(\omega > \omega_B) + \alpha_-(\omega). \tag{B5}$$

The real and imaginary parts of the electric polarizability can be expressed as:

$$\mathrm{Re}[\alpha(\omega > \omega_B)] = \mathrm{Re}[\alpha_+(\omega > \omega_B)] + \alpha_-(\omega), \tag{B6}$$

and

$$\mathrm{Im}[\alpha(\omega > \omega_B)] = \mathrm{Im}[\alpha_+(\omega > \omega_B)]. \tag{B7}$$

**APPENDIX C:** $\alpha_{+1}(\omega < \omega_B)$

We write $\alpha_{+1}(\omega < \omega_B)$ as:

$$\alpha_{+1}(\omega < \omega_B) = \frac{2q^2 e^{2k_o a}}{\omega'}\left(N'(k_0 a)\right)^2\left[\left(\alpha_{+1}(\omega < \omega_B)\right)_I + \left(\alpha_{+1}(\omega < \omega_B)\right)_{II} + \left(\alpha_{+1}(\omega < \omega_B)\right)_{III}\right], \tag{C1}$$

where $N'(k_0 a) = \sqrt{\frac{m}{\hbar^2}}\sqrt{\frac{2k_0}{e^{2k_0 a} + 2k_0 a + 1}}$ , $\omega' = \frac{m}{\hbar}\omega$ and $\left(\alpha_{+1}(\omega < \omega_B)\right)_I$ is given by:

$$(\alpha_{+1}(\omega < \omega_B))_I = \frac{-q^2 k_0}{16\omega'}[-\frac{4a^2 k_0}{\gamma} + 6a^2 - \frac{2a^2\gamma^2}{k_0^2} + \frac{1}{4\omega'^2}(-5k_0^2 - 15\gamma^2 + 16\gamma k_0 + \frac{5\gamma^4}{k_0^2} - \frac{\gamma^6}{k_0^4})], \tag{C2}$$

$\left(\alpha_{+1}(\omega < \omega_B)\right)_{II}$ is given by:

$$(\alpha_{+1}(\omega < \omega_B))_{II} = \frac{q^2 k_0 e^{-2k_0 a}}{24}[\frac{-(4a^3k_0^3 + 16a^2k_0^2 + 3)}{k_0^4} - \frac{1}{k_0^2\omega'}(5a^2\gamma^2 + a^2k_0^2 + 4ak_0 + 3) - \frac{a}{k_0^3\omega'}(5k_0^2 - 3\gamma^2) + \frac{3}{\omega'^3}(k_0^2 + \gamma^2)]. \tag{C3}$$

$\left(\alpha_{+1}(\omega < \omega_B)\right)_{III}$ is given by:

$$(\alpha_{+1}(\omega < \omega_B))_{III} = \frac{q^2 k_0 e^{-2\gamma a}}{4\omega'}[\frac{-a^2 k_0}{\gamma} + \frac{2ak_0}{\omega'} - \frac{\gamma k_0}{\omega'^2}], \tag{C4}$$

where $\gamma^2 = k_0^2 - 2\omega'$.

**APPENDIX D:** $\alpha_{+2}(\omega < \omega_B)$

We write $\alpha_{+2}(\omega < \omega_B)$ as:

$$\alpha_{+2}(\omega < \omega_B) = 8q^2 g' e^{2k_o a}\left(N'(k_0 a)\right)^2 N(\gamma a)\left(\alpha_{+2}(\omega < \omega_B)\right)_I, \tag{D1}$$

where $g' = \frac{m}{\hbar^2} g$ and $N(\gamma a)$ is given by:

$$N(\gamma a) = \frac{1}{\left[1 + \frac{g'}{\gamma}\left(e^{-2a\gamma} - 1\right)\right]} \quad , \tag{D2}$$

and $(\alpha_{+2}(\omega < \omega_B))_I$ is given by:

$$(\alpha_{+2}(\omega < \omega_B))_I = \frac{q^2}{16\omega'^2}[\frac{ak_0}{\gamma} - a - \frac{ak_0 e^{-2a\gamma}}{\gamma} - ae^{-2ak_0} + \frac{k_0 e^{-2a\gamma}}{\omega'} - \frac{k_0 e^{-2ak_0}}{\omega'}]^2 \cdot \tag{D3}$$

**APPENDIX E:** $\alpha_{+1}(\omega' > \omega_B)$

We write $\alpha_{+1}(\omega > \omega_B)$ as:

$$\alpha_{+1}(\omega > \omega_B) = \frac{2q^2 e^{2k_o a}}{\omega'}\left[N'(k_0 a)\right]^2 \begin{bmatrix} (\alpha_{+1}(\omega > \omega_B))_I + (\alpha_{+1}(\omega > \omega_B))_{II} \\ + (\alpha_{+1}(\omega > \omega_B))_{III} + i(\alpha_{+1}(\omega > \omega_B))_{IV} \end{bmatrix}, \tag{E1}$$

where $(\alpha_{+1}(\omega > \omega_B))_I$ is given by:

$$(\alpha_{+1}(\omega > \omega_B))_I = \frac{-q^2 k_0}{16\omega'}\left[\frac{2a^2(3k_0^2 + \Omega^2)}{k_0^2} + \frac{1}{4\omega'^2}(-5k_0^2 + 15\Omega^2 + \frac{5\Omega^4}{k_0^2} + \frac{\Omega^6}{k_0^4})\right] \cdot \tag{E2}$$

$(\alpha_{+1}(\omega > \omega_B))_{II}$ is given by:

$$(\alpha_{+1}(\omega > \omega_B))_{II} = \frac{q^2 k_0 e^{-2k_0 a}}{24}\begin{bmatrix} -\frac{(4a^3k_0^3 + 16a^2k_0^2 + 3)}{k_0^4} + \frac{1}{k_0^2\omega'}(5a^2\Omega^2 - a^2k_0^2 - 4ak_0 - 3) \\ -\frac{a}{k_0^3\omega'}(5k_0^2 + 3\Omega^2) + \frac{3}{\omega'^3}(k_0^2 - \Omega^2) \end{bmatrix} \cdot \tag{E3}$$

$(\alpha_{+1}(\omega > \omega_B))_{III}$ is given by:

$$(\alpha_{+1}(\omega > \omega_B))_{III} = \frac{q^2 k_0}{4\omega'}[\frac{a^2 k_0 Sin(2a\Omega)}{\Omega} + \frac{2ak_0 Cos(2\Omega a)}{\omega'} - \frac{\Omega k_0 Sin(2a\Omega)}{\omega'^2}] \,. \tag{E4}$$

$(\alpha_{+1}(\omega > \omega_B))_{IV}$ is given by:

$$(\alpha_{+1}(\omega > \omega_B))_{IV} = \frac{q^2 k_0}{4\omega'}\left[\frac{a^2 k_0}{\Omega} + \frac{\Omega k_0}{\omega'^2}\right], \tag{E5}$$

where $\Omega^2 = -\gamma^2$.

**APPENDIX F:** $\alpha_{+2}(\omega > \omega_B)$

We write $\alpha_{+2}(\omega > \omega_B)$ as:

$$\alpha_{+2}(\omega > \omega_B) = \frac{q^2 g' k_0^2 e^{2k_o a}}{2\omega'^2 B}\left[N'(k_0 a)\right]^2\left[(\alpha_{+2}(\omega > \omega_B))_I + \mathrm{i}(\alpha_{+2}(\omega > \omega_B))_{II}\right], \tag{F1}$$

$(\alpha_{+2}(\omega > \omega_B))_I$ is given by:

$$(\alpha_{+2}(\omega > \omega_B))_I = \Omega(\Omega - g' Sin(2\Omega a))\left(A^2 - \frac{a^2}{\Omega^2}\right) - 2aAg' \cdot \tag{F2}$$

$(\alpha_{+2}(\omega > \omega_B))_{II}$ is given by:

$$(\alpha_{+2}(\omega > \omega_B))_{II} = 2aA\left(\Omega - g'Sin(2\Omega a)\right) + \Omega g'\left(A^2 - \frac{a^2}{\Omega^2}\right) , \tag{F3}$$

where

$$A = \frac{-a}{k_0} + \frac{aSin(2a\Omega)}{\Omega} - \frac{ae^{-2k_0a}}{k_0} + \frac{Cos(2a\Omega)}{\omega'} - \frac{e^{-2k_0a}}{\omega'} , \tag{F4}$$

and

$$B = \left(\Omega - g'Sin(2\Omega a)\right)^2 + (g')^2 . \tag{F5}$$

**APPENDIX G: α(0)**

α(0) is given by:

$$\alpha(0) = \frac{q^2 m}{\hbar^2}\left[\frac{a^2\left(N(k_0a)\right)^2}{k_0^3}\left[\frac{-(3+16ak_0)}{6} + \frac{ge^{4k_0a}}{\left[g + e^{2k_0a}(-g + k_0)\right]}\right] + \frac{\left(5 + 12a^2k_0^2\right)}{4k_0^4}\right]. \tag{G1}$$

[a)] Email: anis.maize@email.stvincent.edu
[b)] Email: john.smetanka@email.stvincent.edu